\documentclass{elsart5p}

\usepackage{graphics}
\usepackage{graphicx}
\usepackage{amssymb}
\begin{document}

\begin{frontmatter}


\title{\textbf{Large magnetocaloric effect in $\mbox{Ho}_2\mbox{Pd}_{2}\mbox{Pb}$}}

\author{Baidyanath Sahu}, 
\author{R. Djoumessi Fobasso}, 
\author{Buyisiwe M Sondezi}, and 
\author{Andr\'{e} M. Strydom}

\address{Highly Correlated Matter Research Group, Department of Physics, University of Johannesburg, PO Box 524, Auckland Park 2006, South Africa}

\corauth[abd]{Corresponding author. baidyanathsahu@email.com}

\begin{abstract}

We report the magnetocaloric effect (MCE) in a polycrystalline plumbide sample of $\mathrm{Ho_2Pd_2Pb}$. Arc-melted $\mathrm{Ho_2Pd_2Pb}$ crystallizes in $\mathrm{Mo_2B_2Fe}$ type of tetragonal crystal structure and shows an antiferromagnetic behavior with N\'{e}el temperature of 4.5 K. The Arrott plots indicate that $\mathrm{Ho_2Pd_2Pb}$ undergoes a second order antiferromagnetic phase transition. The calculated isothermal magnetic entropy change ($\Delta S_m$), and relative power cooling (refrigeration capacity) for a change of field 0-8 T are 17.8 J.kg$^{-1}$.K$^{-1}$, and 680(497)~J.kg$^{-1}$ respectively. The obtained results revealed that $\mathrm{Ho_2Pd_2Pb}$ belongs to a family of large MCE magnetic materials. 

\end{abstract}


\end{frontmatter}


\section{INTRODUCTION}

The magnetocaloric effect (MCE) is an intrinsic property of magnetic materials that manifests itself by an adiabatic variation of temperature ($\Delta T_{ad}$) or the isothermal variation of magnetic entropy ($\Delta S_m$) when subjected to a magnetic field. This effect is maximal around the  magnetic phase transition temperature. At this point, the temperature rate of change of the magnetization of the material undergoing the magnetic ordering is maximum and results in a strong variation of magnetic entropy, which due to the magneto-thermal coupling induces a variation of temperature of the material. Based on the magnetocaloric effect, magnetic refrigeration is a less polluting technology compared to conventional refrigeration (absence of harmful gases, better energy efficiency). The higher energy efficiency and environmentally friendly nature of this technique has motivated the search for magnetic materials that exhibit large MCE and are suitable for potential applications. The large/giant MCE materials can be classified from the values of $\Delta S_m$, $\Delta T_{ad}$ and refrigeration capacity (RC)/relative power cooling (RCP) \cite{book1, book2,book3,book, RTX}. 

The $\mathrm{R_2T_2X}$ series of compounds (R = rare-earth elements, T= transition metal, X = general elements) have attracted much attention to their crystal structure, and related physical properties. Relatively large MCE values have already been found in some $\mathrm{R_2T_2X}$ compounds \cite{R2T2X}. The crystal structures of tetragonal $\mathrm{Mo_2B_2Fe}$$\textendash$ type systems together with exploratory magnetic characterizations have already been reported for $\mathrm{R_2Pd_2Pb}$ compounds \cite{R2Pd2Pb, RE2Pd2Pb}. In this paper, the large MCE found in an arc-melted polycrystalline compound of $\mathrm{Ho_2Pd_2Pb}$ is reported.


\section{EXPERIMENTAL DETAILS}
A polycrystalline sample of $\mathrm{Ho_2Pd_2Pb}$ was prepared on a water$\textendash$cooled copper hearth by high purity Argon arc$\textendash$melting method with a tungsten electrode. 
To promote homogeneity the ingot was turned over and remelted several times. 
Powder X-ray diffraction (XRD) at room temperature was carried out using a Cu-K$\alpha$ radiation ($\lambda$ = 1.5405~{\AA}) on a Rigaku diffractometer (SmartLab).  The phase purity and crystal structure analysis were performed by refining the powder XRD patterns using Rietveld refinement with FullProf software \cite{rietveld, fullprof}. The temperature and field dependence of the dc-magnetization measurements were carried out using a Dynacool Physical Property Measurement system (PPMS) (made by Quantum Design, San Diego, USA) with a Vibrating Sample Magnetometer (VSM) option, in the temperature range of 2-300 K, with the dc-magnetic field ($H$) from 0 to 9 T. Temperature dependence of magnetization ($M(T)$) was measured under zero field cooled (ZFC) and field cooled (FC) protocol \cite{Gd2Rh3Ge}. The magnetic entropy change was calculated from the isothermal field$\textendash$dependent  magnetization curves using the Maxwell relation \cite{book}:
\begin{equation}
\Delta S_m (T, H) = \mu_0 \int_{H_i}^{H_f} \left(\frac{\partial M}{\partial T}\right)_{H'} dH'
\label{DeltaS}
\end{equation}
where $H_i$ is the initial magnetic field and $H_f$ is the final field.

\section{RESULTS AND DISCUSSIONS}

\begin{figure}[!t]
	\includegraphics[scale=0.35]{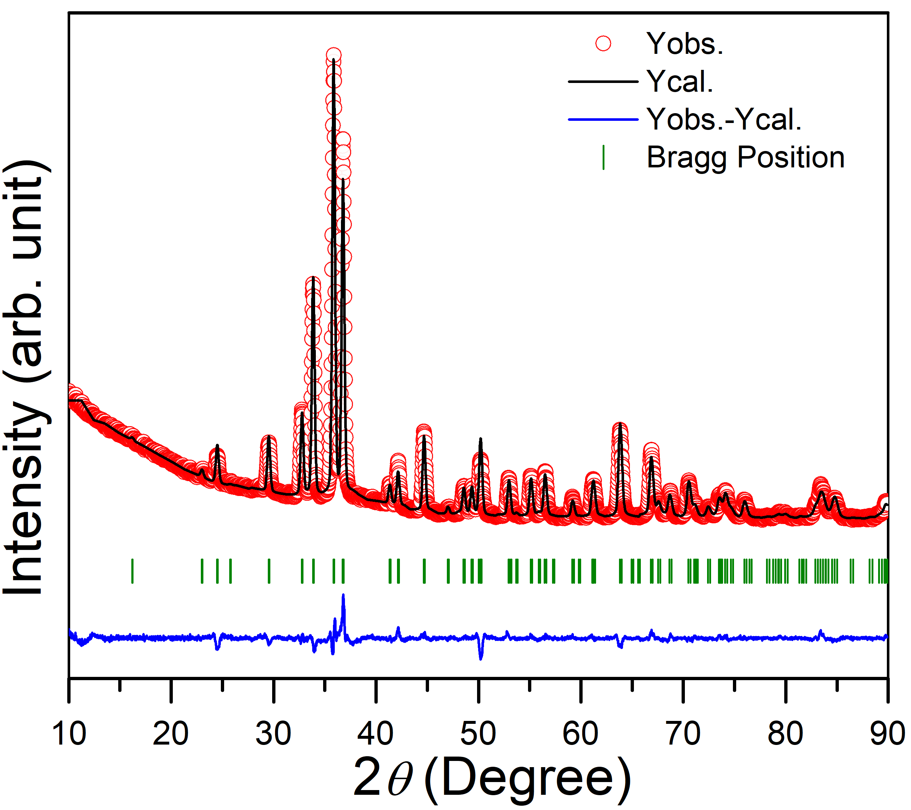}
	\caption{Rietveld refinement fitting on the room temperature powder XRD pattern of $\mathrm{Ho_2Pd_2Pb}$.}
	\label{XRD}
\end{figure}

\begin{figure}[!t]
	\includegraphics[scale=0.35]{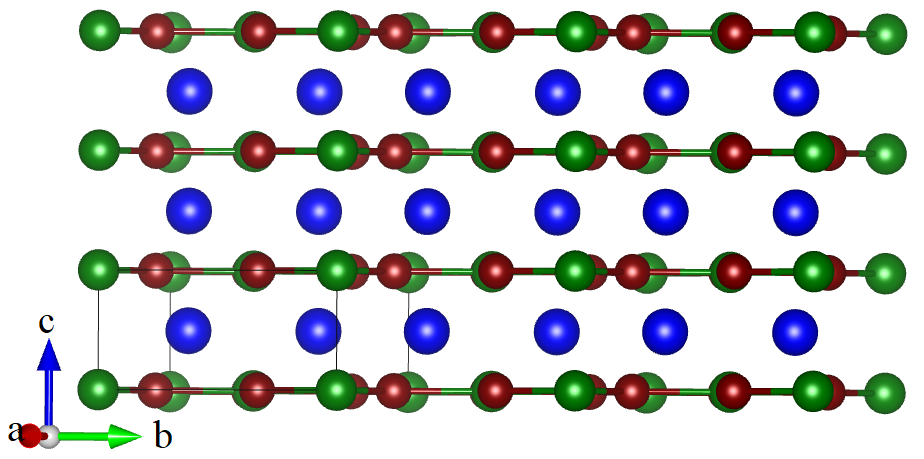}
	\caption{Layered structure of $\mathrm{Ho_2Pd_2Pb}$, and blue, magenta, and green colour balls are for Ho, Pd and Pb respectively.}
	\label{stucture}
\end{figure}

Fig.~\ref{XRD} shows the Rietveld refinement of recorded room temperature XRD patterns reveals that $\mathrm{Ho_2Pd_2Pb}$ crystallizes in a $\mathrm{Mo_2B_2Fe}$$\textendash$type of tetragonal structure with space group $P4/mbm$. The yielded lattice parameters from the refinement are a~=~b~=~7.7282(2)\AA{} and c~=~ 3.6327(1) \AA{}. These values are in good agreement with the previously reported values \cite{R2Pd2Pb}. The full description of crystal structure and interatomic distance is refereed to the published paper \cite{R2Pd2Pb}. Here, Fig~\ref{stucture} displays the arrangement of atoms in layers in the b-c plane of tetragonal structure  for $\mathrm{Ho_2Pd_2Pb}$. One layer contains the rare-earth element Ho, and other layer contains both Pd and Pb elements.

\begin{figure}[!t]
	\includegraphics[scale=0.295]{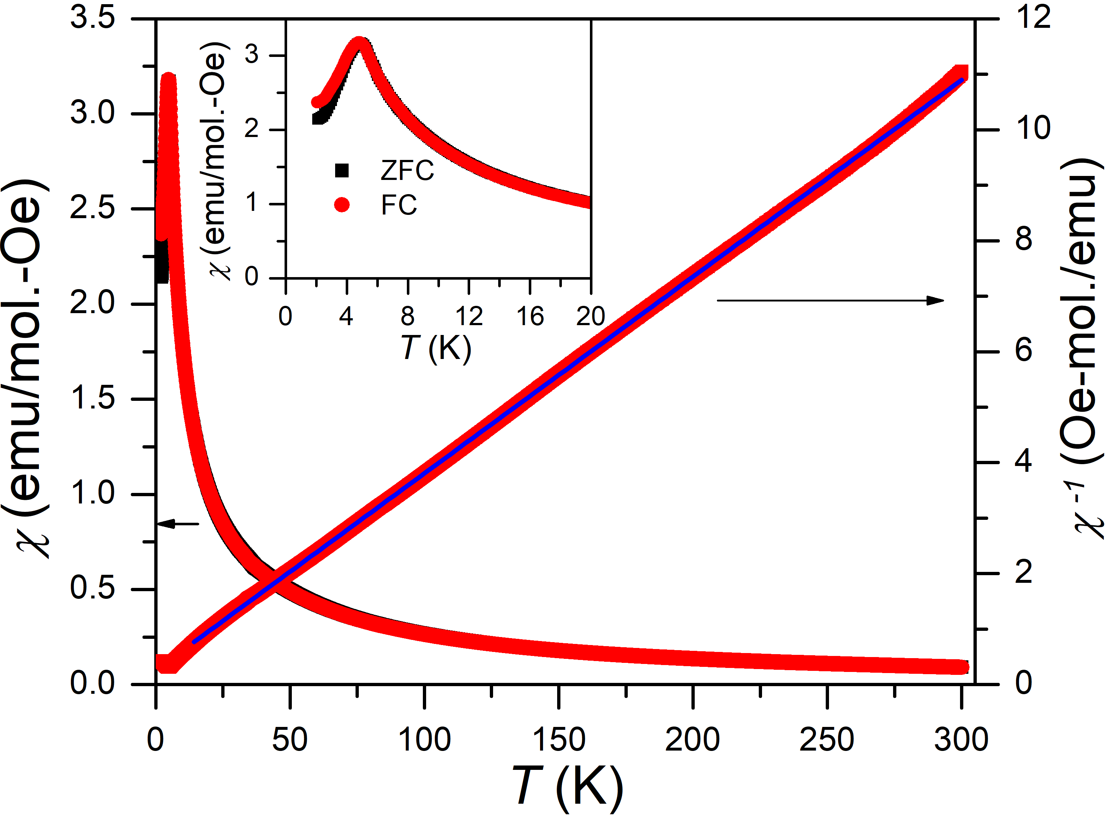}
	\caption{left scale; Temperature dependence of ZFC and FC dc$\textendash$magnetic susceptibility of $\mathrm{Ho_2Pd_2Pb}$, and right scale; Inverse dc$\textendash$magnetic susceptibility in 0.1 T of FC curve along with the Curie$\textendash$Weiss fitting. Inset: Expanded low temperature region of dc$\textendash$magnetic susceptibility .}
	\label{MT}
\end{figure}

The temperature dependence of ZFC and FC dc$\textendash$magnetic susceptibility ($\chi(T)$ = $M(T)/H$) (left scale) and the inverse magnetic susceptibility $\chi^{-1}(T)$ of FC curve (right scale) under a magnetic field $H$ = 0.05 T for $\mathrm{Ho_2Pd_2Pb}$ are shown in Fig.~\ref{MT}. $\chi(T)$ results proved that $\mathrm{Ho_2Pd_2Pb}$ undergoes an antiferromagnetic to paramagnetic transition with N\'{e}el temperature $T_N$ = 4.5 K,  which was estimated from the maximum values in ZFC and FC of $\chi(T)$ curves and shown in the inset of Fig.~\ref{MT}. It is observed that the FC and ZFC curves almost completely overlap, indicating no thermomagnetic irreversibility in  $\mathrm{Ho_2Pd_2Pb}$. The linear paramagnetic region above 10 K of the $\chi^{-1}(T)$ follows Curie–Weiss-type behaviour; given by the relation $\chi = N_{\mathrm{A}} \mu_{\mathrm{eff}}^2/3k_{\mathrm{B}}(T - \theta_{\mathrm{P}})$, 
with $ N_{\mathrm{A}}$ the Avogadro number and $k_{\mathrm{B}}$ the Boltzmann constant. 
The least-squares fit yields the effective magnetic moment $\mu_{\mathrm{eff}}$~=~10.62~ $\mu_{\mathrm{B}}$/Ho. This obtained $\mathrm{\mu_{eff}}$ is very close to the theoretical value of a free $\mathrm{Ho^{3+}}$ ion, $\mathrm{g_{J}[J(J + 1)]^{1/2}}$ = 10.61 $\mathrm{\mu_{B}}$ for J = 8, and g = 5/4. This result indicates that the 4f$\textendash$shell electrons of $\mathrm{Ho^{3+}}$ ions are the predominant magnetic species in $\mathrm{Ho_2Pd_2Pb}$. The negative Weiss paramagnetic temperature $\theta_{\mathrm{P}} = -7.4 $ K suggests the predominance of antiferromagnetic interactions in the compound. The results are in good agreement with the previous report \cite{RE2Pd2Pb}.

\begin{figure}[!t]
	\includegraphics[scale=0.31]{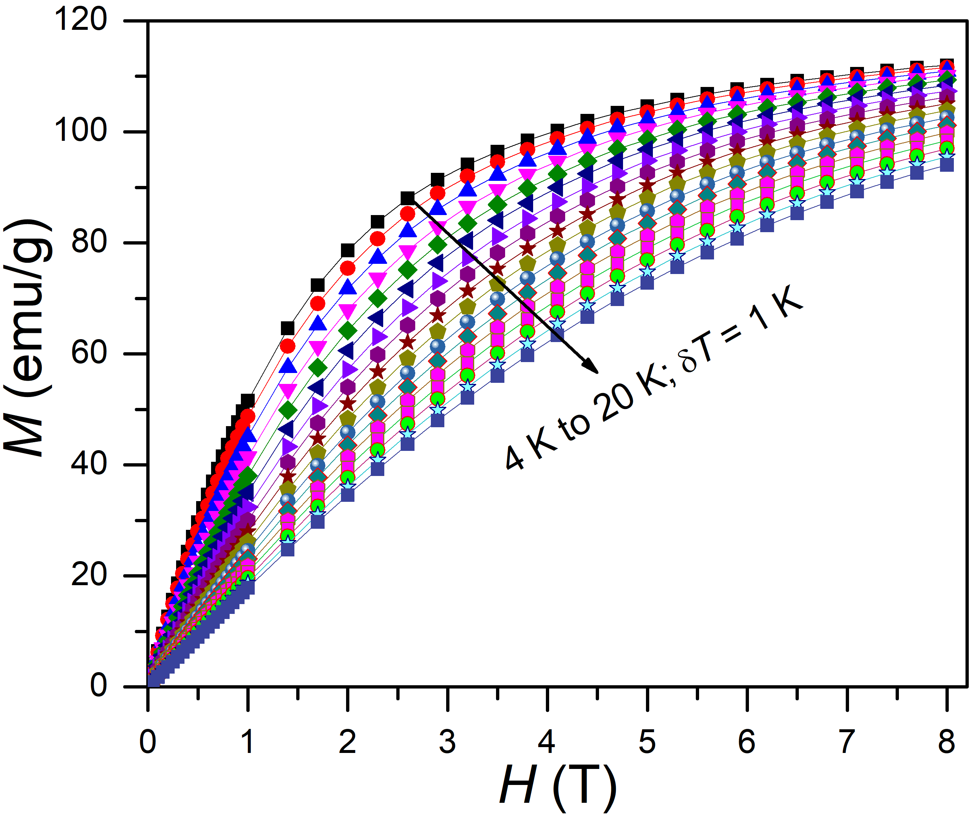}
	\caption{Isothermal magnetization at different temperatures ranging from 4 K to 20 K with a step of 1 K for $\mathrm{Ho_2Pd_2Pb}$.}
	\label{MH}
\end{figure}

\begin{figure}[!t]
	\includegraphics[scale=0.31]{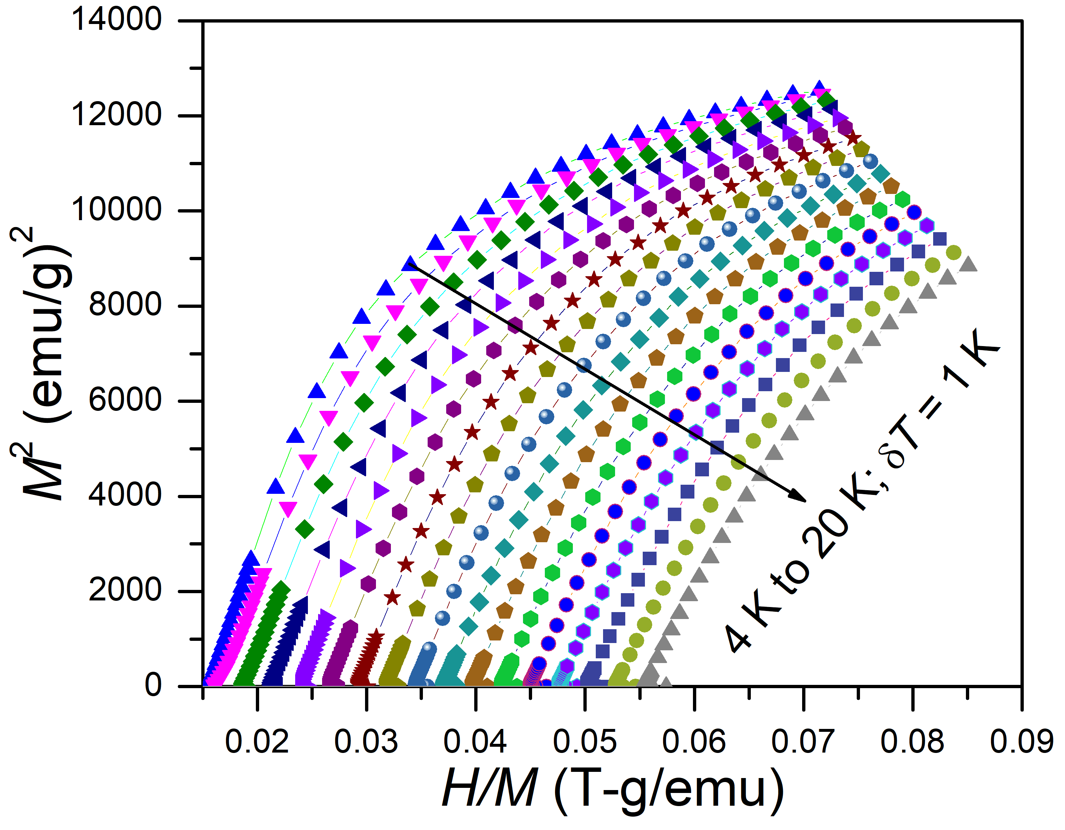}
	\caption{Arrott plots, $M^2~vs.~H/M$ at different temperature from 4 K to 20 K with a step of 1 K for $\mathrm{Ho_2Pd_2Pb}$.}
	\label{AP}
\end{figure}

To evaluate the MCE of $\mathrm{Ho_2Pd_2Pb}$, magnetic isothermal $M(H)$ is measured in a wide temperature range from 4 to 20 K under applied magnetic field up to 8 T and is shown in Fig.~\ref{MH}. To avoid field cycling effect between consecutive isothermal magnetization measurements, the sample was heated to 50 K between each magnetization isotherm. Hysteresis loop is not seen in $M(H)$ below $T_N$. In order to determine the order of the magnetic phase transition of $\mathrm{Ho_2Pd_2Pb}$,  $M$$^2$~$vs.$~$H/M$ plots are made for various temperatures from $M(H)$ and is shown in Fig.~\ref{AP}. According to the Banerjee criterion \cite{Banerjee}, the positive slopes of Arrott plots represent a second$\textendash$ order magnetic transition and negative slopes at some point represent first$\textendash$order transitions. As seen in Fig.~\ref{AP}, the positive slopes in Arrott plots across the entire $M(H)$ range are characteristic of the second$\textendash$order magnetic phase transition in $\mathrm{Ho_2Pd_2Pb}$.

\begin{figure}[!t]
	\includegraphics[scale=0.31]{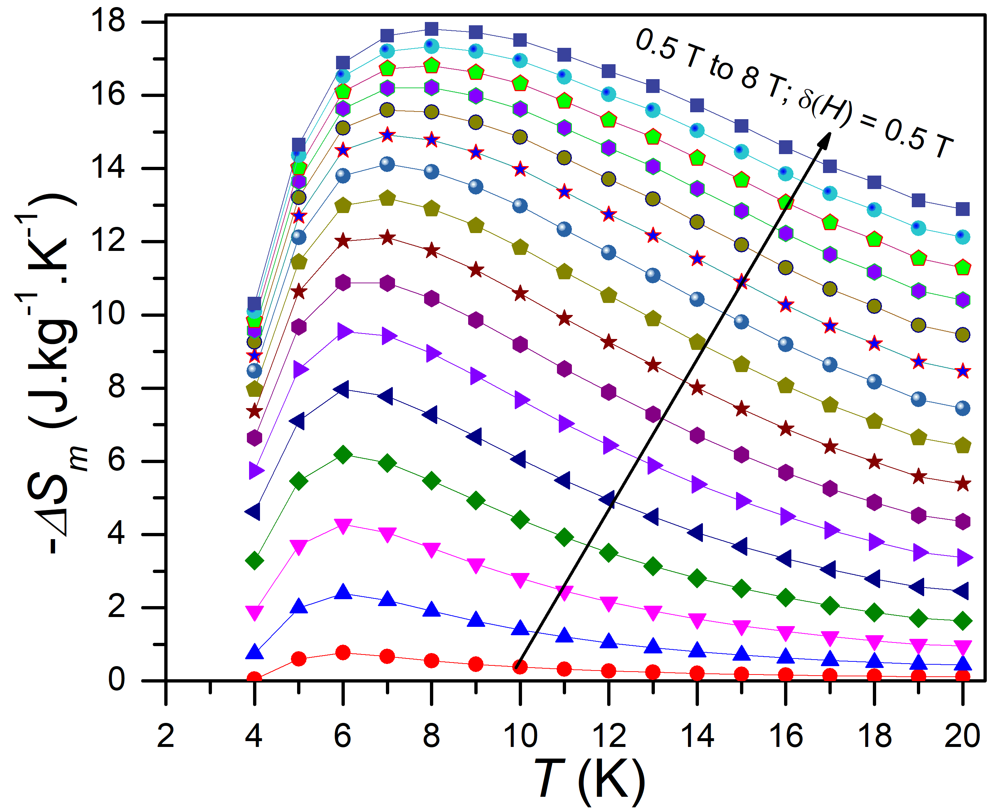}
	\caption{The isothermal magnetic entropy change as a function of temperature for various
	magnetic field changes from 0$\textendash$8 T for $\mathrm{Ho_2Pd_2Pb}$.}
	\label{MCE}
\end{figure}

Magnetic entropy changes were derived from  $M(H)$, using  Maxwell's thermodynamic relation given in equation~\ref{DeltaS}. Fig.~\ref{MCE} shows the temperature dependence of $- \Delta S_m$ at different magnetic fields changing up to 9 T. It is observed that $- \Delta S_m (T,H)$ exhibits a broad peak around the transition temperature.  The maximum values of the magnetic entropy change ($- \Delta S^{max}_m$) for $\mathrm{Ho_2Pd_2Pb}$ are found to be 6.2 , 14.1, 16.8 and 17.8 $\mathrm{J.kg^{-1}.K^{-1}}$ for the change of magnetic field 0-2 T, 0-5 T, 0-7 T and 0-8 T, respectively.

The RCP and RC  are also considered as important figures of merit to quantify the heat transfer between the hot and cold reservoirs in an ideal
refrigeration cycle. The RCP is defined as the product of the maximum magnetic entropy change $|- \Delta S^{max}_m|$ and the full width at half maximum ($\delta T_{FMHM})$ of  $\Delta S_m$~$vs.$~$T$ curves; RCP~$=$~$|- \Delta S^{max}_m|$~$\times$~$\delta T_{FMHM}$ \cite{book, RTX, RCP, RCP1}. The RC is determined by the formula: $RC~=~\int_{T_1}^{T_2} |\Delta S_m| dT$. Here, $T_1$ and $T_2$ are the temperatures below and above $T_C$, respectively, at the half maximum of the $\Delta S_m$ peak in an ideal thermodynamic cycle. The RC and RCP values obtained using the above formulas gradually increase with applied magnetic fields. The RCP(RC) values are 61(43), 270(190), 470(340) and 605(435) J/kg for the change of field 0$\textendash$2, 0$\textendash$5, 0$\textendash$7 and 0$\textendash$8 T respectively. A table is made to compare the MCE performance of $\mathrm{Ho_2Pd_2Pb}$ with other $\mathrm{Ho_2T_2X}$ MCE materials. As seen from the table, the $- \Delta S^{max}_m$ value of $\mathrm{Ho_2Pd_2Pb}$ is comparable with the reported values of other Ho$\textendash$based 2:2:1 compounds. However, RCP value is smaller than the reported ferromagnetic compounds and larger than the antiferromagnetic magnetic compounds. This comparison suggests that the $\mathrm{Ho_2Pd_2Pb}$ belong to a class of large MCE material.

\begin{table}[!t]
\caption{\label{tableMCE} The transition temperature (($T_M$) and * represents  Curie temperature for ferromagnetic transition), the magnetic entropy change ($\Delta S_m$), and relative cooling power (RCP) under a magnetic field change of 0$\textendash$5~T for $\mathrm{Ho_2Pd_2Pb}$ together with reported MCE of $\mathrm{Ho_2T_2X}$ compounds.}
\begin{tabular}{ccccc} \\ \hline
\hspace{-0.01 in}Compound	 & 	\hspace{0.17 in} $T_M$	 & 	\hspace{0.17 in}$-\Delta S_m$	 &	 \hspace{0.17 in}RCP	 &		 \hspace{0.17 in}Ref  \\

\hspace{-0.01 in} & \hspace{0.17 in} (K) & \hspace{0.17 in} (J/kg-K) & \hspace{0.17 in} (J kg$^{-1}$) & \hspace{0.17 in}   \\ \hline	

\hspace{-0.01 in}$\mathrm{Ho_2Cu_2Cd}$ & \hspace{0.17 in} 30/15* & \hspace{0.17 in} 20.3 & \hspace{0.17 in} 481 & \hspace{0.17 in} \cite{Ho2Cu2Cd}  \\

\hspace{-0.01 in}$\mathrm{Ho_2Cu_2In}$ & \hspace{0.17 in} 30* & \hspace{0.17 in} 17.4 & \hspace{0.17 in} 416 & \hspace{0.17 in} \cite{Ho2Au2In}  \\

\hspace{-0.01 in}$\mathrm{Ho_2Au_2In}$ & \hspace{0.17 in} 21/8* & \hspace{0.17 in} 12.9 & \hspace{0.17 in} 345 & \hspace{0.17 in} \cite{Ho2Au2In}  \\

\hspace{-0.01 in}$\mathrm{Ho_2Co_2Ga}$ & \hspace{0.17 in} 38.5* & \hspace{0.17 in} 11.7 & \hspace{0.17 in} 271 & \hspace{0.17 in} \cite{Ho2Co2Ga}  \\

\hspace{-0.01 in}$\mathrm{Ho_2Co_2Al}$ & \hspace{0.17 in} 27/9* & \hspace{0.17 in} 11.5 & \hspace{0.17 in} 580 & \hspace{0.17 in} \cite{Ho2Co2Al}  \\

\hspace{-0.01 in}$\mathrm{Ho_2Ni_2Ga}$ & \hspace{0.17 in} 12.5 & \hspace{0.17 in} 5.4 & \hspace{0.17 in} 121 & \hspace{0.17 in} \cite{Ho2Ni2Ga}  \\

\hspace{-0.01 in}$\mathrm{Ho_2Ni_2In}$ & \hspace{0.17 in} 10.5/5.5 & \hspace{0.17 in} 11.5 & \hspace{0.17 in} 210 & \hspace{0.17 in} \cite{Ho2Ni2In}  \\

\hspace{-0.01 in}$\mathrm{Ho_2Ni_2Al}$ & \hspace{0.17 in} 12/6 & \hspace{0.17 in} 6 & \hspace{0.17 in} 151 & \hspace{0.17 in} \cite{Ho2Co2Al}  \\

\hspace{-0.01 in}$\mathrm{Ho_2Pd_2Pb}$ & \hspace{0.17 in} 4.5 & \hspace{0.17 in} 14.1 & \hspace{0.17 in} 270 & \hspace{0.17 in} PW  \\
\hline
PW: Present Work 
\end{tabular}	
\end{table}

\section{SUMMARY}

In summary, a single$\textendash$phase $\mathrm{Ho_2Pd_2Pb}$ compound with $\mathrm{Mo_2B_2Fe}$-type tetragonal
structure was successfully synthesized by arc-melting technique. $\mathrm{Ho_2Pd_2Pb}$ exhibits a second order antiferromagnetic phase transition with a N\'{e}el temperature of 4.5 K. The estimated maximum values of $- \Delta S_m$ and RCP (RC) around the transition temperature of $\mathrm{Ho_2Pd_2Pb}$  are 17.8 $\mathrm{J.kg^{-1}.K^{-1}}$ and 605(435) J.kg$^{-1}$, respectively  for a change of magnetic field 0-8 T. These results may provide some interest to search for new $\mathrm{R_2T_2X}$ series of MCE materials.

\section*{ACKNOWLEDGEMENT}   
This work is supported by Global Excellence and Stature (UJ-GES) fellowship, University of Johannesburg, South Africa. DFR thanks OWSD and SIDA for the fellowship towards PhD studies. BMS thanks the URC (283153) of UJ for financial support. AMS thanks the URC/FRC of UJ and SA NRF (93549) for financial assistance.

\subsection*{\label{sec:level3}Data availability}
The data that support the findings of this study are available from the corresponding author upon reasonable request.

\section*{Author contributions}
This manuscript has been written through the contributions of
all authors. All authors have given approval to the final version of the manuscript.

\section*{Conflict of interest}
All authors declare that there is no conflict of financial interests or personal relationships.

\end{document}